\documentclass[a4paper,11pt]{article}
\pdfoutput=1 

\usepackage{jheppub} 

\usepackage[T1]{fontenc} 
\usepackage{subfig}
\usepackage{mathtools}
\usepackage{float}
\usepackage{xcolor}
\usepackage{soul}
\usepackage{placeins}

\title{\boldmath Preserving physically important variables in optimal event selections: A case study in Higgs physics}

\newcommand{\met}{E^\text{miss}_\text{T}}
\newcommand{\pet}{p^\text{miss}_\text{T}}
\newcommand{\mbb}{m_{bb}}
\newcommand{\mbbi}{(m_{bb})_i}
\newcommand{\ttbar}{t\bar{t}}
\newcommand{\mi}{\text{MI}(\hat{y}, \mbb)}

\newcommand{\T}{T_{\theta_a}[\hat{y}_i, \mbbi]}

\author[a]{Philipp Windischhofer}
\author[a]{Miha Zgubi\v{c}}
\author[a]{Daniela Bortoletto}

\affiliation[a]{Department of Physics, University of Oxford,\\ Oxford, United Kingdom}

\emailAdd{philipp.windischhofer@physics.ox.ac.uk}
\emailAdd{miha.zgubic@physics.ox.ac.uk}

\abstract{Analyses of collider data, often assisted by modern Machine Learning methods, condense a number of observables into a few powerful discriminants for the separation of the targeted signal process from the contributing backgrounds. These discriminants are highly correlated with important physical observables; using them in the event selection thus leads to the distortion of physically relevant distributions.
We present a novel method based on a differentiable estimate of mutual information, a measure of non-linear dependency between variables, to construct a discriminant that is statistically independent of a number of selected observables, and so manages to preserve their distributions in the event selection. Our strategy is evaluated in a realistic setting, the analysis of the Standard Model Higgs boson decaying into a pair of bottom quarks. Using the distribution of the invariant mass of the di-$b$-jet system to extract the Higgs boson signal strength, our method achieves state-of-the-art performance compared to other decorrelation techniques, while significantly improving the sensitivity of a similar, cut-based, analysis published by ATLAS.
}

\begin{document} 
\maketitle
\flushbottom

\section{Introduction}
The discovery of a Standard Model (SM)-like Higgs boson by the ATLAS \cite{Aad:2012tfa} and CMS \cite{Chatrchyan:2012xdj} collaborations gave rise to a rich experimental programme, seeking to measure its properties with the highest degree of precision. The Higgs boson has since been observed in its main production modes and a range of decay channels \cite{Aaboud:2018urx, Sirunyan:2017khh, ATLAS:2014aga, Khachatryan:2016vau, ATLAS-CONF-2018-031, Sirunyan:2018kst, Aaboud:2018zhk}.

Many of these results rely on machine learning algorithms at various stages of the analysis pipeline, most prominently for the separation of signal and background processes. Often the entire analysis is designed around a powerful multivariate discriminant computed by a boosted decision tree (BDT) or artificial neural network. The Higgs boson signal strength is commonly extracted through a maximum-likelihood fit to the distribution of this discriminant \cite{Aaboud:2018urx, Aaboud:2018zhk}.

As shown in e.g.~\cite{Cranmer:2016arx}, such a discriminant approximates a sufficient statistic of the
data, and can therefore exploit the discriminating power and the correlations between input variables in the most optimal way. While statistically
very intuitive, it can however not be directly related to individual physical observables. More importantly, its
sufficiency drains all other event variables of their discriminative power to separate signal and background: once the
discriminant is used to select a sample of signal-enriched events, even the distributions of physically highly relevant
observables (such as the invariant mass of a system of particles) get distorted and lose their conventional interpretations.

In the past, this has led to the development of dedicated ``cross-check'' analyses. Typically, these do not strive to maximise the expected signal sensitivity, but rather try to directly operate in terms of individual physical observables.
For example, the observation of the Higgs boson decay to two bottom quarks, $H\rightarrow b\bar{b}$, by the
ATLAS and CMS collaborations was complemented by such an alternative analysis \cite{Aaboud:2018zhk, Sirunyan:2018kst}. Its event selection is purely based on simple cuts on a few physically relevant event variables. The Higgs boson signal strength is then extracted through a maximum-likelihood fit to the invariant mass of the two $b$-jets, $m_{bb}$, arguably the most physically relevant variable in this setting and a quantity whose meaning is easily conveyed.

The performance of such an analysis is governed by the ability of the event selection to separate signal from background, subject to the constraint that the distortion of the $m_{bb}$ distribution of the background be limited. Only then can the Higgs boson signal strength and the background normalisation be simultaneously constrained by the fit.
A similar problem has recently appeared in jet tagging, where increasingly powerful machine learning-based algorithms are seen to distort the spectrum of the jet mass. A lot of progress has been made in the construction of taggers whose outputs are \textsl{decorrelated} with (more precisely, statistically independent of) the mass. Such taggers therefore limit the extent to which such distortions occur upon the selection of a signal-enriched subset of jets \cite{ATLAS_decorrelated, decorrelated_substructure, DisCo}.

In this letter, we introduce a novel method for the training of such a decorrelated classifier, based on the adversarial minimisation of a neural estimate of mutual information. We apply our approach to the analysis of the process mentioned earlier, $H\rightarrow b\bar{b}$. In particular, we target the production of the Higgs boson in association with an electroweak vector boson ($V = W$, $Z$) and focus on the 0-lepton channel of its decay. The analysis of this process is complicated by the presence of a number of challenging backgrounds. Following the cut-based analysis of this process published by ATLAS \cite{Aaboud:2018zhk}, we design an event classification procedure that separates events into multiple regions, while seeking to preserve the $m_{bb}$ distributions in each category, and using this variable as the final discriminant in terms of which the fit is performed.

In particular, the aim of this approach is not---and cannot be---to improve upon the sensitivity and precision delivered by a traditional multivariate analysis. Instead, our goal is to evaluate the feasibility of this alternative analysis strategy, benchmark our approach against other decorrelation strategies proposed in the literature \cite{Louppe:2016ylz, DisCo} and the ATLAS cut-based analysis, and assess its performance in a realistic and complementary setting.

The remainder of this letter is structured as follows. Section \ref{simulation} describes the generation of the dataset on which our studies are based. Then, Section \ref{methodology} introduces our method for decorrelation and the analysis in which it is embedded. Section \ref{results} presents our findings and Section \ref{conclusions} provides an outlook to future work.

\section{Simulation Techniques and Validation}
\label{simulation}
Monte Carlo (MC) simulation is used to model the signal and background processes at a centre-of-mass energy of $\sqrt{s}=13$\,TeV. The signal sample contains a Higgs boson $H$ produced in association with a vector boson $V$ and subsequently decaying to a pair of bottom quarks, $VH\rightarrow b\bar{b}$. The main background processes are $W$ and $Z$ production in association with $b$-jets ($W$+jets, $Z$+jets), top quark pair ($\ttbar$) production, and diboson production.

For all samples, the hard scattering is generated using \textsc{MadGraph5} 2.6.5 \cite{Alwall:2014hca}, while parton showering is implemented using \textsc{Pythia} 8.2 \cite{Sjostrand:2007gs}. The inclusive cross-sections are normalised to the NNLO calculation where available and to NLO otherwise \cite{Vjets_xsec, ttbar_xsec}. The yields are further scaled to correspond to the full LHC Run II integrated luminosity of $140~\text{fb}^{-1}$. The $\ttbar$ background is simulated at next-to-leading order (NLO) accuracy, while a leading-order (LO) calculation was found to be sufficient for the other processes. The detector response is simulated using the \textsc{Delphes} 3.4.1 \cite{deFavereau:2013fsa} parametrisation of the ATLAS detector.

After performing the event selection outlined in Section \ref{event_selection}, we validate our setup by comparing the per-process event yields against the numbers shown in Table 13 of \cite{Aaboud:2017xsd}. The signal yields were found to agree within 5\%, while individual background yields were found to agree within about 20\%. The largest deviation in the yields was observed for the $\ttbar$ process. Our dataset reproduces the overall background normalisation to within 8\%.

The effect of pileup collisions on the implemented event selection was studied as well. Given the information made available by \textsc{Delphes}, we found it difficult to accurately reproduce the performance of the ATLAS pileup suppression techniques. Considering that these effects are of subleading importance to a relative comparison of different methods, contributions from pileup are switched off in our dataset.

\section{Analysis Methodology}
\label{methodology}
Section \ref{event_selection} describes the common event selection used for all subsequent studies. The cut-based analysis employed by ATLAS is described in Section \ref{cba} as a baseline approach. Section \ref{pca} describes the proposed alternative event selection and classification scheme employing a decorrelated classifier.

\subsection{Event selection}
\label{event_selection}
The event selection closely follows that of Ref.~\cite{Aaboud:2018zhk}. The analysis considers only events containing exactly two $b$-jets, and one additional non-$b$-flavoured jet is permitted. To make efficient use of the generated MC samples, this $b$-tagging requirement is implemented at the level of the generated partons. Detector effects are then included by multiplying the event weight by the $b$-tagging efficiencies achieved by the individual $b$-jets, according to \cite{ATL-PHYS-PUB-2015-022}. Only events without reconstructed leptons (electrons or muons) with a transverse momentum of $p_T>7$\,GeV are used.

The leading $b$-jet is required to have a transverse momentum larger than 45\,GeV and the missing transverse energy in the event ($\met$) is required to be larger than 150\,GeV.
A requirement on the scalar sum of the jet $p_T$, $H_\text{T} > 120~(150)$\,GeV for events containing 2 (3) jets, is
used to remove phase space where trigger efficiencies depend on the jet multiplicity. Cleaning cuts on the angular
separation between the jets, and between the di-$b$-jet system and $\met$ are used to remove events in which the $\met$
results from mismeasured jets. A requirement based on $\Delta \phi (\met, \pet)$, used in the ATLAS analysis for the same purpose,
is dropped\footnote{$\pet$ is the missing transverse momentum of tracks in the inner detector matched to the primary
vertex.}.

\subsection{Cut-based analysis}
\label{cba}
The ATLAS cut-based analysis (CBA) \cite{Aaboud:2018zhk} makes use of four signal regions. The events passing the above selection are first divided into two categories, containing events with two and three jets, respectively. These two categories are further split into two regions each, containing events with $150 < \met < 200$\,GeV (low-$\met$ categories) and $\met > 200$\,GeV (high-$\met$ categories). An additional cut on the separation of the two $b$-jets, $\Delta R(b_1, b_2) < 1.8~(1.2)$ is applied in the low (high)-$\met$ categories in order to suppress events with large $\mbb$\footnote{$\Delta R(b_1, b_2) = \sqrt{\Delta \phi^2 + \Delta \eta^2}$ where $\Delta \phi$ and $\Delta \eta$ are the differences in the azimuthal angle and pseudorapidity between the two $b$-jets.}.

\begin{figure*}[tp]
  \centering
  \subfloat[][]{\includegraphics[width=0.48\textwidth]{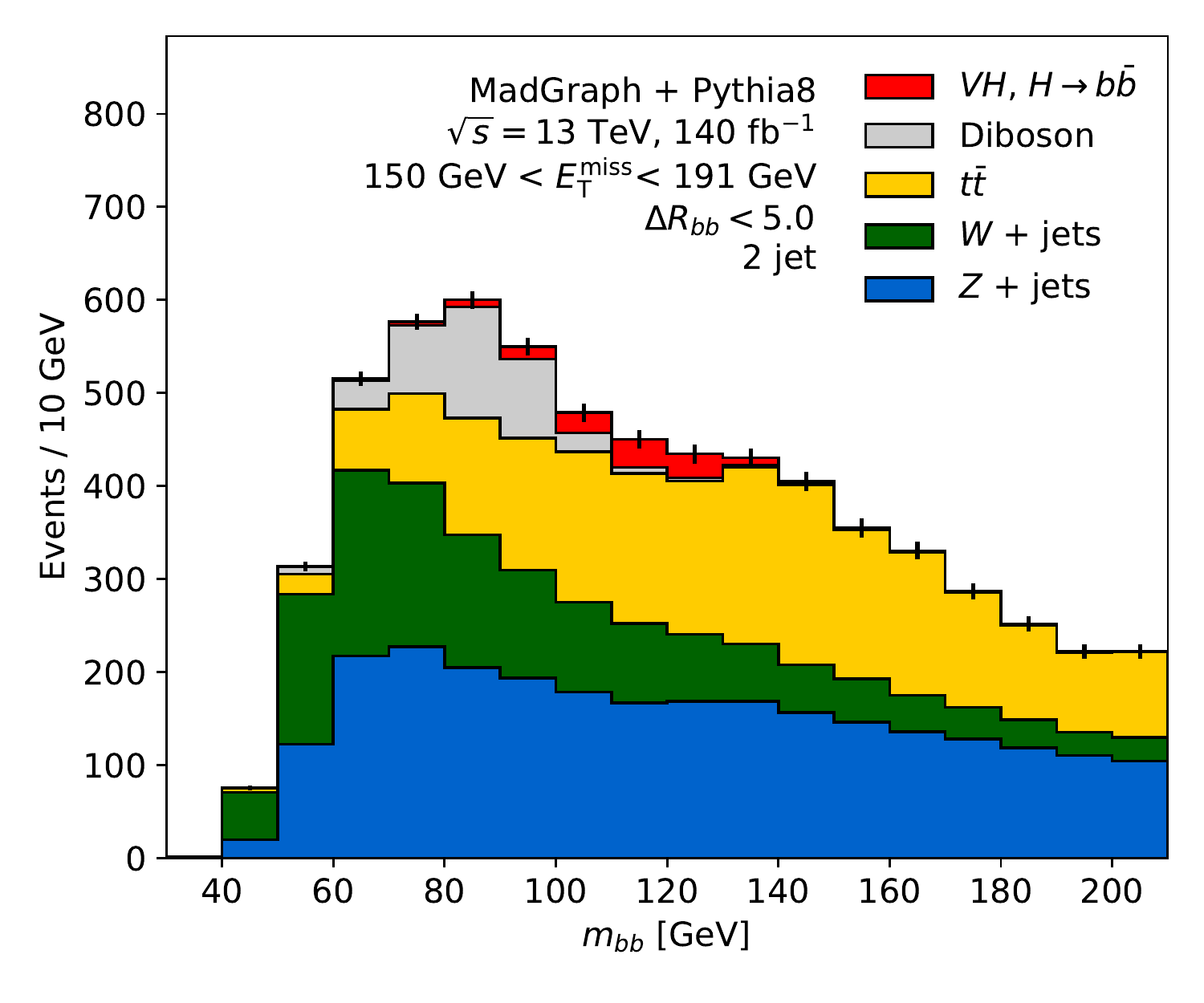}}
  \subfloat[][]{\includegraphics[width=0.48\textwidth]{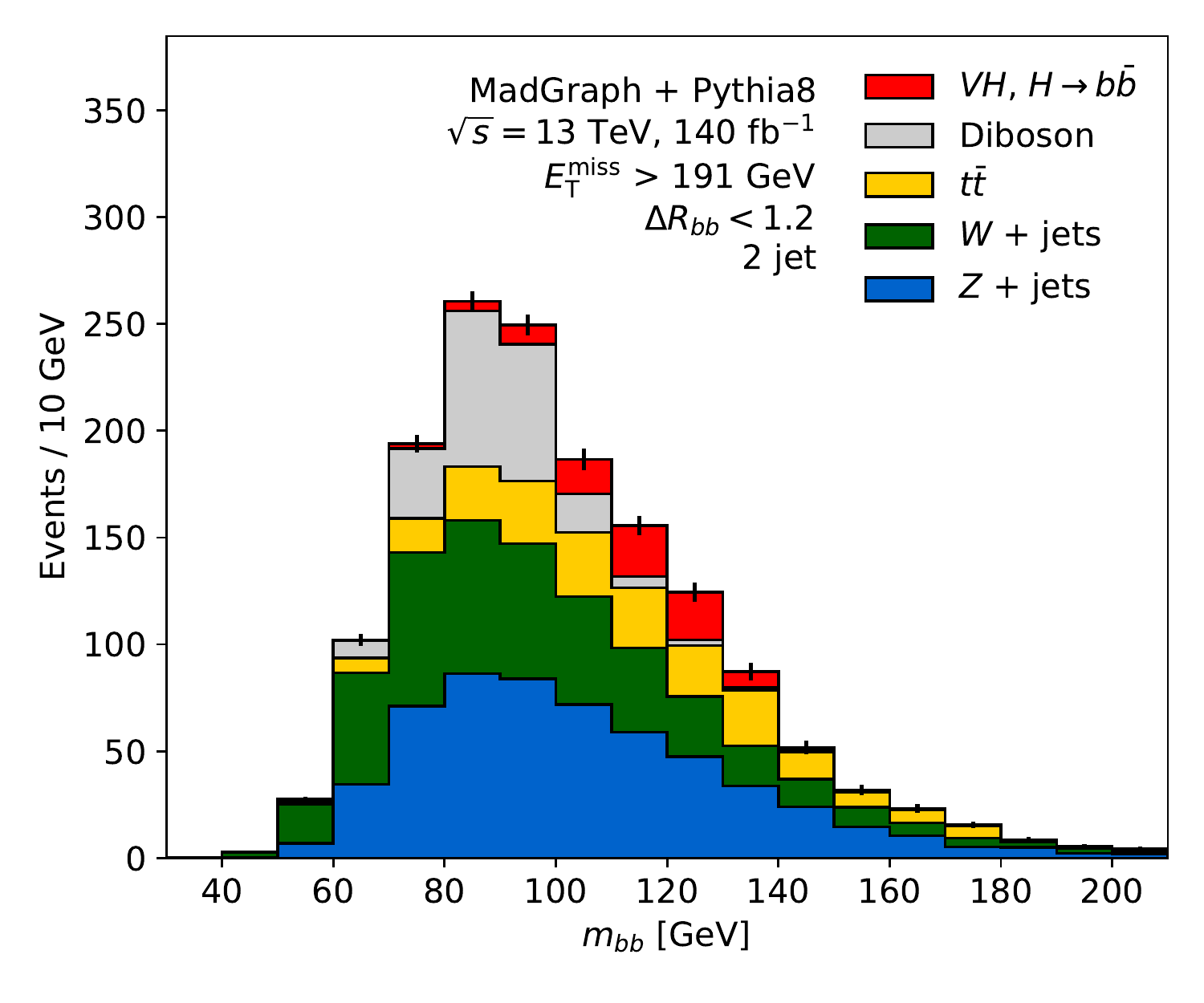}}\\
  \subfloat[][]{\includegraphics[width=0.48\textwidth]{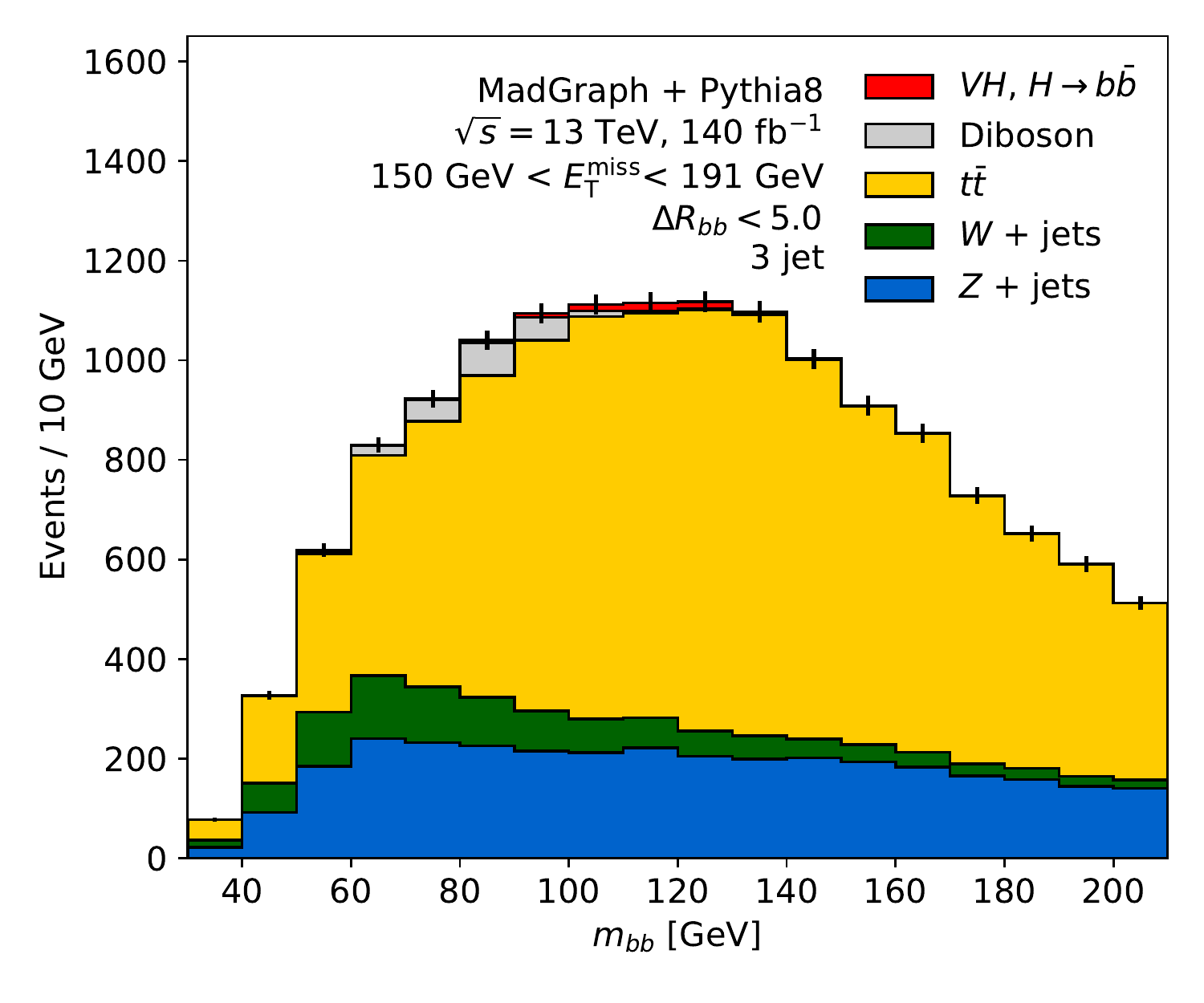}}
  \subfloat[][]{\includegraphics[width=0.48\textwidth]{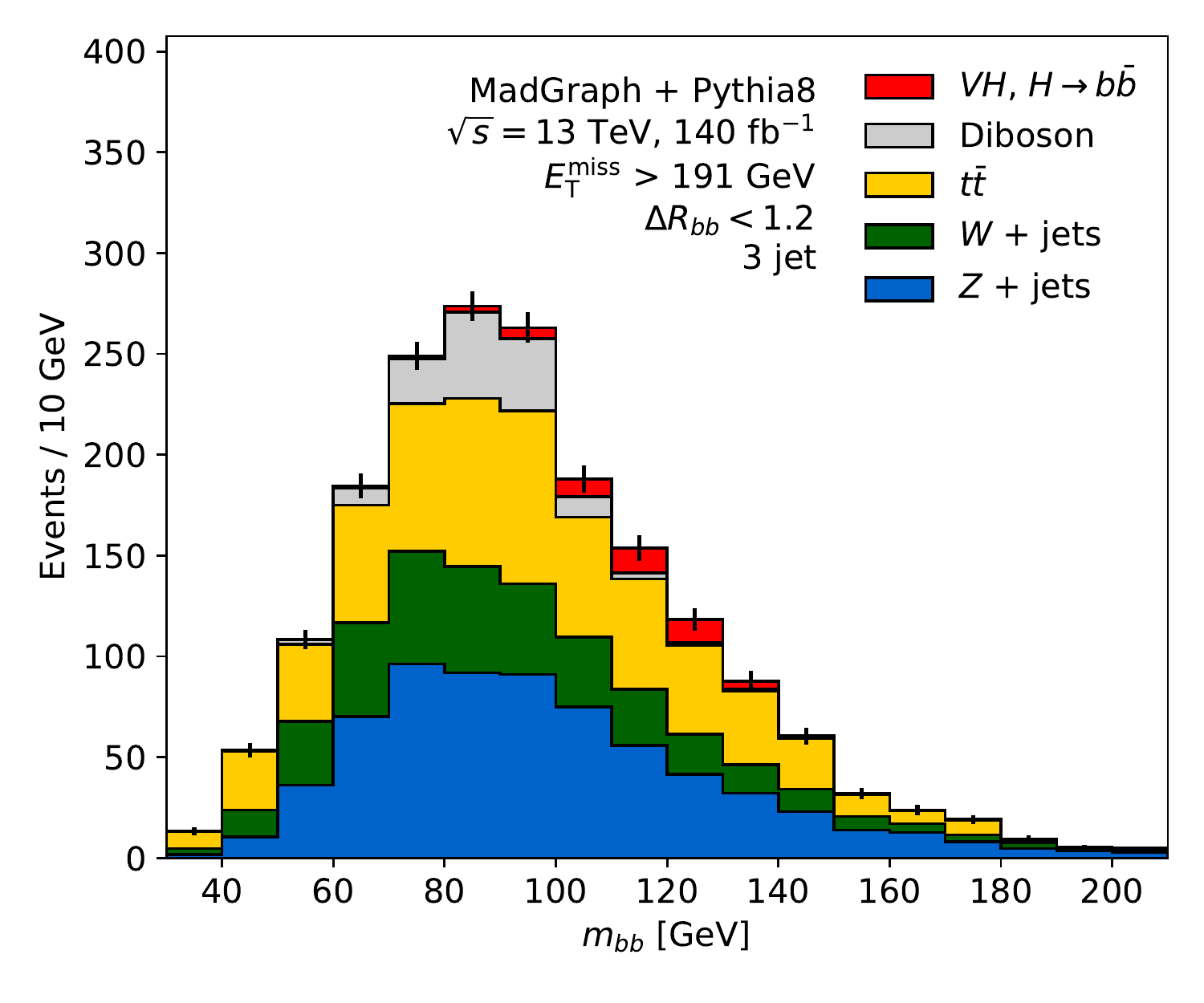}}\\
  \caption{CBA categories. The $\mbb$ distribution is shown in the two-jet low-$\met$ (a) and high-$\met$ (b)
  categories, as well as in the three-jet low-$\met$ (c) and high-$\met$ (d) categories. Signal events are shown in red, while
  backgrounds are shown in blue ($Z$+jets), green ($W$+jets), yellow ($\ttbar$), and gray (diboson).
  Error bars show the statistical uncertainty on the total MC prediction and are dominated by the uncertainty on the background.}
  \label{fig:cba}
\end{figure*}

Given the simplifications made in the generation of our dataset and the observed residual yield differences, these cut values are not necessarily optimal. They are thus re-optimised to maximise the sensitivity of the CBA on our MC samples, using Bayesian optimisation \cite{gp}. The optimised cuts are chosen to maximise the Asimov significance of the Higgs boson signal, as obtained from a simultaneous maximum-likelihood fit to all analysis regions (as defined in Section \ref{results}).
This procedure results in a selection of $150 < \met < 191$\,GeV ($\met > 191$\,GeV) and $\Delta R(b_1, b_2) < 5.0~(1.2)$ for the low (high)-$\met$ regions, respectively. Distributions of the di-$b$-jet invariant mass $\mbb$ in each of the CBA signal regions defined by these optimised cuts are shown in Figure \ref{fig:cba}.

\subsection{Decorrelated classifier analysis}
\label{pca}
Our proposed analysis strategy is based on the same underlying event selection. The event categories are now defined through cuts on a multivariate discriminant $\hat{y}$ computed by a neural network. It is trained to separate signal from background, i.e.~solves a classification problem.

To avoid any distortion of the $\mbb$ distributions of the background upon the selection of signal-like events, $\hat{y}$ must be \emph{statistically independent} from $\mbb$, i.e.~the joint probability distribution $p(\mbb, \hat{y})$ of the background must factorise,
\begin{equation}
  p(\mbb, \hat{y}) = p(\mbb)\cdot p(\hat{y}) \implies p(\mbb|\hat{y}) = p(\mbb).
  \label{independence}
\end{equation}
In slightly less precise terms, the classifier output $\hat{y}$ is decorrelated with $m_{bb}$. We thus refer to this analysis strategy as \emph{decorrelated classifier analysis} (DCA).

The requirement of statistical independence is incorporated through a modification of the loss function that is
used to train the classifier neural network \cite{Louppe:2016ylz, Zhang2018MitigatingUB}. We add an auxiliary loss term $\mathcal{L}_{\text{aux}}$ to the usual classification loss $\mathcal{L}_{\text{clf}}$, suitably constructed to impose the independence constraint in Equation \ref{independence},
\begin{equation}
  \mathcal{L} = \mathcal{L}_{\text{clf}} + \lambda\cdot \mathcal{L}_{\text{aux}}.
  \label{global_loss}
\end{equation}
Here, $\lambda$ is a continuous (but fixed) parameter that compromises between the (generally conflicting) objectives specified by $\mathcal{L}_{\text{aux}}$ and $\mathcal{L}_{\text{clf}}$. Below, we introduce the form of $\mathcal{L}_{\text{aux}}$ for our mutual information-based method. Other decorrelation methods proposed in the literature are briefly reviewed in Section \ref{other_decorrelation_techniques}.

\subsubsection{Mutual information-based decorrelation}
\label{sec:MIND}
The mutual information (MI) between the classifier response and the invariant mass, $\mi$, is defined as the Kullback-Leibler divergence \cite{kullback1951} between the joint and the product of the marginal probability densities,
\begin{equation}
  \mi = D_\text{KL}\left(p{(\hat{y}, \mbb)} || p{(\hat{y})}\, p{(\mbb)}\right) = \int \text{d}\mbb\, \text{d}\hat{y}\,\, p{(\hat{y}, \mbb)}\cdot\log{\left(\frac{p{(\hat{y}, \mbb)}}{p{(\hat{y})} \cdot p{(\mbb)}}\right)}.
  \label{MI_definition}
\end{equation}
This object is a measure of (nonlinear) relationship between the two random variables: $\mi$ is zero if and only if $\hat{y}$ and $\mbb$ are statistically independent, and positive otherwise. Equation \ref{MI_definition} expresses MI as a nonlinear functional of the joint density $p{(\hat{y}, \mbb)}$ and its marginals, a fact that causes much of the difficulty of estimating MI. Many estimators of MI have been developed in the past \cite{doi:10.1162/089976603321780272, cellucci_binning_heuristics}. However, most of these techniques are not directly applicable to our setting, as they either require a binning of the data (such as those using histograms for density estimation) or are not straightforwardly applicable to weighted data (when using a $k$-nearest neighbours estimate of the densities). Most importantly, none of these estimators are differentiable functions of their inputs, and so cannot be used in conjunction with gradient descent-based optimisation, as we require here.

As shown in \cite{nguyen2010estimating, nowozin2016f}, the KL divergence---more generally, any $f$-divergence---admits a dual representation that allows to rephrase the estimation of MI from a set of samples as a functional optimisation problem.
This makes the estimation of MI tractable with modern numerical optimisation techniques. Indeed, MI can alternatively be defined as
\begin{equation}
  \mathrm{MI}(\hat{y}, m_{bb}) = \sup_{T_0}\,\, \langle T_0 \rangle_{p(\hat{y}, m_{bb})} - \log\left( \langle e^{T_0} \rangle_{p(\hat{y})p(m_{bb})} \right),
  \label{MINE_def}
\end{equation}

where the supremum is taken over all functions $T_0: (\hat{y}, m_{bb}) \mapsto \mathbb{R}$ such that both expectations are finite. The form of Equation \ref{MINE_def} allows MI to be approximated with arbitrary accuracy from a batch of $N_b$ samples drawn from $p(\hat{y}, m_{bb})$. Indeed, as \cite{belghazi2018mine} shows, $\widehat{\mathrm{MI}}^{\mathrm{MINE}}$, a strongly consistent estimator of MI, can be obtained by replacing the expectation values in Equation \ref{MINE_def} with the equivalent sample averages,
\begin{equation}
  \widehat{\mathrm{MI}}^{\mathrm{MINE}}(\hat{y}, m_{bb}) := \sup_{\theta_a}\,\, \frac{1}{N_b} \left[ \sum^{N_b}_{\substack{i = 0\\p{(\hat{y}, \mbb)}}} \T - \log\left( \sum^{N_b}_{\substack{i=0\\p{(\hat{y})}\,
      p{(\mbb)}}} e^{\T} \right) \right].
  \label{MINE_estimator}
\end{equation}
In the first sum $\hat{y}$ and $\mbb$ are sampled from their joint probability density while in the second they are sampled from the respective marginals. The function $T_{\theta_a}$ is parametrised by an ancillary ``adversary'' neural network with weights $\theta_a$. The adversary network is trained to extremise the objective on the right-hand side of Equation \ref{MINE_estimator} through standard gradient-based methods. It thus plays an integral role in defining $\mathcal{L}_{\text{aux}}$ which regularises the classifier loss $\mathcal{L}_{\text{clf}}$.

The estimate $\widehat{\mathrm{MI}}_{\mathrm{MINE}}$ does not require the samples to be binned and is a manifestly differentiable function of the inputs. Thus, it can simply be used in the auxiliary loss term in Equation \ref{global_loss},
\begin{equation}
  \mathcal{L}_{\text{aux}} = \widehat{\mathrm{MI}}^{\mathrm{MINE}}_{\mathrm{bkg}},
  \label{MINE_loss}
\end{equation}
thus forcing the classifier output to become statistically independent of $m_{bb}$. The subscript makes explicit that the independence requirement is only imposed on the background, i.e.~the sums in Equation \ref{MINE_estimator} are restricted to background events.

The application discussed here requires the estimation of MI of two one-dimensional random variables. However, our method is not limited to this case and generalises straightforwardly to higher-dimensional random vectors. Provided that a sufficiently expressive architecture is chosen for the adversarial neural network, the classifier can be decorrelated with multiple event variables, which can be discrete or continuous.

Our method thus explicitly minimises \textbf{M}utual information to make the classifier statistically \textbf{Ind}ependent of $m_{bb}$. Below we refer to it as \textsl{MIND}.

\subsubsection{Other decorrelation strategies}
\label{other_decorrelation_techniques}
Other decorrelation techniques have been proposed in the literature and some have been applied in a particle physics context. The methods we compare to MIND are briefly summarised in the following\footnote{This is by far not an exhaustive collection of methods, but instead focuses on methods that are immediately applicable to our situation. A simple reweighting of signal and background to have the same $m_{bb}$ distribution is not feasible here because of the very peaked structure of the Higgs signal, leading to event weights of very large magnitude. We also investigated a simple adversarial approach where the adversary acts as a regressor, trying to infer $m_{bb}$ from $\hat{y}$. However, this approach proved to be too weak and could not achieve any substantial degree of decorrelation.}. For further details, references to the original publications are given.

\paragraph{Explicit entropy maximisation (EMAX)}
Introduced in \cite{Louppe:2016ylz}, this strategy maximises an estimate for the conditional entropy between $m_{bb}$ and $\hat{y}$ for background events,
\begin{equation}
  \mathcal{L}_{\mathrm{aux}} = -\hat{\mathrm{H}}_{\mathrm{bkg}}(m_{bb}|\hat{y}) = \langle \log p(m_{bb}|\hat{y}) \rangle_{p(\hat{y}, m_{bb})}.
  \label{GMM_loss}
\end{equation}

By virtue of the relation
\begin{equation}
  \mathrm{MI}(\hat{y}, m_{bb}) = \mathrm{H}(m_{bb}) - \mathrm{H}(m_{bb}|\hat{y}),
\end{equation}
this leads to a minimisation of MI, and thus makes the classifier output $\hat{y}$ independent of MI.

In practice, this approach requires the choice of an explicit parametric model for $p(m_{bb}|\hat{y})$. In our application, a Gaussian mixture model was found to be sufficient. For more complicated situations, or in a higher-dimensional setting, normalising flows \cite{normalising_flows} might be a viable alternative. If we denote the free parameters of this model as $\theta_a$ and the modelled density as $p_{\theta_a}(m_{bb}|\hat{y})$, the conditional entropy estimator used in Equation \ref{GMM_loss} is (using the same notation as in Section \ref{sec:MIND})
\begin{equation}
  -\hat{\mathrm{H}}_{\mathrm{bkg}}(m_{bb}|\hat{y}) = \sup_{\theta_a} \frac{1}{N_b} \sum_{\substack{i = 0\\p(\hat{y}, m_{bb})}}^{N_b} \log p_{\theta_a}(\left(m_{bb}\right)_i|\hat{y}_i).
  \label{GMM_aux_loss}
\end{equation}

\paragraph{DisCo}
While this paper was under review, a new method was published in \cite{DisCo}. Conceptually very similar to our strategy, distance covariance between $\hat{y}$ and $m_{bb}$, $\mathrm{dCov}(\hat{y}, m_{bb})$, is used to rate the degree of (non-linear) dependence between these variables. It is defined as the (suitably weighted) $L^2$ distance between the joint characteristic function $\phi_{(\hat{y}, m_{bb})}$ and the product of its marginals,
\begin{equation}
  \mathrm{dCov}(\hat{y}, m_{bb}) := || \phi_{(\hat{y}, m_{bb})} - \phi_{(\hat{y})} \phi_{(m_{bb})} ||_2^w
  \label{dcov_definition}
\end{equation}
where $w$ denotes the weighting function. Instead of relying on Equation \ref{dcov_definition} directly, \cite{DisCo} suggest to use a scaled variant, the distance correlation,
\begin{equation}
  \mathrm{dCorr}(\hat{y}, m_{bb})^2 = \frac{\mathrm{dCov}^2(\hat{y}, m_{bb})}{\mathrm{dCov}(\hat{y}, \hat{y})\, \mathrm{dCov}(m_{bb}, m_{bb})}.
\end{equation}
Similar to MI, distance correlation is nonnegative and zero if and only if $\hat{y}$ and $m_{bb}$ are independent.

Notably, a differentiable sample estimate $\widehat{\mathrm{dCorr}}$ of distance correlation can be constructed explicitly (see \cite{DisCo} for the full expression) without the need for a dedicated adversarial network, and so the auxiliary loss term is directly given by this estimate, again restricted to the background,
\begin{equation}
  \mathcal{L}_{\mathrm{aux}} = \widehat{\mathrm{dCorr}}^2_{\mathrm{bkg}}(\hat{y}, m_{bb}).
\end{equation}

\subsubsection{Classifier architecture}
\label{classifier_architecture}

We can equivalently formulate the constraint in Equation \ref{independence} as
\begin{equation}
  p(\hat{y}|\mbb) = \int dx\,\, p(\hat{y}|x)\, p(x|\mbb) = p(\hat{y}).
\end{equation}
The object $p(\hat{y}|x_i)$ is the probability density of the classifier response $\hat{y}$ generated for an event with input features $x_i$. It is thus natural to think of the classifier as defining a probability distribution rather than implementing a deterministic function, i.e.~as a \emph{randomised classifier} \cite{Thie:2016rclf}. Generalising the classifier model in this way extends the class of problems for which useful decorrelated discriminants can be constructed \cite{randomised_classifier_fairness}.

We choose to parametrise $p(\hat{y}|x_i)$ through a mixture of $N$ Gaussian distributions, transformed to limit its support to the interval $(0, 1)$. The $3N - 1$ free parameters of this model (the means $\mu_k$, standard deviations $\sigma_k$ as well as the mixture fractions) are computed by the classifier network.
The response $\hat{y}_i$ to an event $x_i$ is then defined as a random variate drawn from this distribution, i.e.~$\hat{y}_i \sim p(\hat{y}|x_i)$. Despite the inherent stochasticity of this classifier architecture, coupling the random seed to a unique identifier (such as the event number) ensures a reproducible event classification.

For training of the randomised classifier, the usual binary cross-entropy loss is used. Explicitly, on a batch of $N_b$ events, the loss takes the form
\begin{equation}
\mathcal{L}_{\text{clf}} = - \frac{1}{N_\text{b}}\sum_{i=1}^{N_\text{b}} y_i \log{\hat{y}_i} + (1 - y_i) \log{(1 - \hat{y}_i)},
\end{equation}
where $\hat{y}_i$ labels the classifier response and $y_i$ represents the true identity of the event, i.e.~whether it originates from signal or background.

The input variables used for the classifier are a subset of those used for the BDTs employed by the main analysis in \cite{Aaboud:2018zhk},
all known to be reliably modelled by MC. These include $\mbb$, $\Delta R_{bb}$, $\Delta \eta_{bb}$, $\met$, the transverse momenta
of the two $b$-jets, the scalar sum of all jet momenta as well as the azimuthal angle
difference between the di-$b$-jet system and $\met$. It is worth noting that the classifier not only has access to
variables highly correlated with $\mbb$, such as $\Delta R_{bb}$, but also to $\mbb$ itself. As shown in Section \ref{results}, the
adversarial training is nevertheless able to reliably achieve independence from the invariant mass.

\subsubsection{Training methodology}
We first describe the training process for the methods using an adversarial network, MIND and EMAX. The training proceeds in two distinct phases \cite{goodfellow2014generative, Louppe:2016ylz}. First, the adversary network is pretrained for $N_p = 1000$ minibatch steps by updating $\theta_a$ such as to compute the suprema in Equations \ref{MINE_loss} and \ref{GMM_aux_loss},
\begin{equation}
  \theta_a \gets \theta_a + \eta_a \nabla_{\theta_a} \mathcal{L}_{\text{aux}}.
  \label{adversary_update}
\end{equation}
For MIND, this provides an initial estimate of $\mi$ of the untrained classifier.

In the second phase, the classifier weights $\theta_c$ are updated, taking into account the auxiliary loss term,
\begin{equation}
  \theta_c \gets \theta_c - \eta_c \nabla_{\theta_c} \left( \mathcal{L}_{\text{clf}} + \lambda\cdot\mathcal{L}_{\text{aux}} \right),
  \label{classifier_update}
\end{equation}
followed by an update of the adversarial network according to Equation \ref{adversary_update} in order to adapt to the changes in the classifier. To allow the adversary to
continuously follow the classifier, a larger learning rate is chosen for the latter, $\eta_a > \eta_c$, and multiple updates are applied to the adversary weights $\theta_a$.

Owing to the different definitions of $\mathcal{L}_{\mathrm{aux}}$, different ranges of the hyperparameter $\lambda$ are relevant for each method. For MIND, this parameter is varied in the range $\lambda_{\mathrm{MIND}} \in (0, 40)$, while $\lambda_{\mathrm{EMAX}}\in (0, 500)$ was chosen for EMAX. The upper limits for $\lambda$ are selected to capture the maximum analysis performance as measured by the Asimov fit significance introduced in Section \ref{results}.

For both MIND and EMAX, the classifier (adversary) is implemented by a fully connected neural network with 3 (6) layers of 30 nodes each, using ReLU activations \cite{ReLU}. For EMAX, the adversary implements a Gaussian mixture model with 10 components.

The Adam \cite{kingma2015adam} optimisation algorithm is used for all studies. The learning rates are set to constant values of $\eta_a = 4\cdot 10^{-4}$ and $\eta_c = 2\cdot 10^{-4}$. The training is carried out for 30k minibatches and a batch size of $N_{b}=1000$ was found to work well in our application. No significant improvements were observed for larger batch sizes or smaller learning rates. Each batch is chosen to contain the same overall contribution from signal and background; the relative normalisation of the various background components is set to correspond to their abundance in the full training set. The training makes use of $1/3$ of the available simulated events (about 800k events both for signal and background), with the remaining events held back for performance validation.

Separate randomised classifiers (with separate adversaries) are trained for events containing two and three jets, sharing the same value of $\lambda$.
Already a single Gaussian was found to be sufficient for the parametrisation of the classifier output distribution
$p(\hat{y}|x)$. Using a deterministic classifier instead was found to yield comparable performance but slightly worse training stability.

For DisCo, the training proceeds in the same way, except that the initial pretraining of the adversarial network and its repeated updates during the training of the classifier are absent. The same classifier architecture was used and $\lambda_{\mathrm{DisCo}}\in (0, 10)$ is selected. The batch size is increased to $N_b = 2000$ ($N_b = 5000$) events for the 2-jet (3-jet) channel to ensure numerical stability. In practice, this leads to a significantly longer training time for DisCo compared to the other methods, even in the absence of a dedicated adversary network.

No selection of the best-performing models is performed after training. Instead, the analysis sensitivity achieved for all trained classifiers is shown in Section \ref{results} in order to communicate the variance in these results that can reasonably be expected for the various decorrelation techniques.

Our implementation of the three considered decorrelation strategies can be accessed at \cite{gitlab}. It makes use of the \texttt{TensorFlow} library and the randomised classifier is based on the \texttt{TensorFlow Probability} package \cite{tensorflow2015-whitepaper}.

\subsubsection{Event categorisation}
Following the training, events are classified based on $\hat{y}$. \textit{Tight} and \textit{loose} regions are defined, separately for events with two and three jets. Events are allocated to the \textit{tight} or \textit{loose} categories based on the ordering defined by the discriminant $\hat{y}$, with the two \textit{tight} categories containing the most signal-like events.

For simplicity, and to ease the comparison with other decorrelation methods and with the CBA, the \textit{loose} (\textit{tight}) category is constructed to achieve the same signal efficiency as the low (high)-$\met$ categories, separately for both jet multiplicities. The signal efficiencies found to be optimal for the CBA are thus taken to be optimal also for the DCA. Potential additional performance improvements coming from a dedicated opimisation of the classifier working points are therefore not realised.

\section{Results}
\label{results}

\begin{figure*}[!!h]
  \centering
  \includegraphics[width=0.72\textwidth]{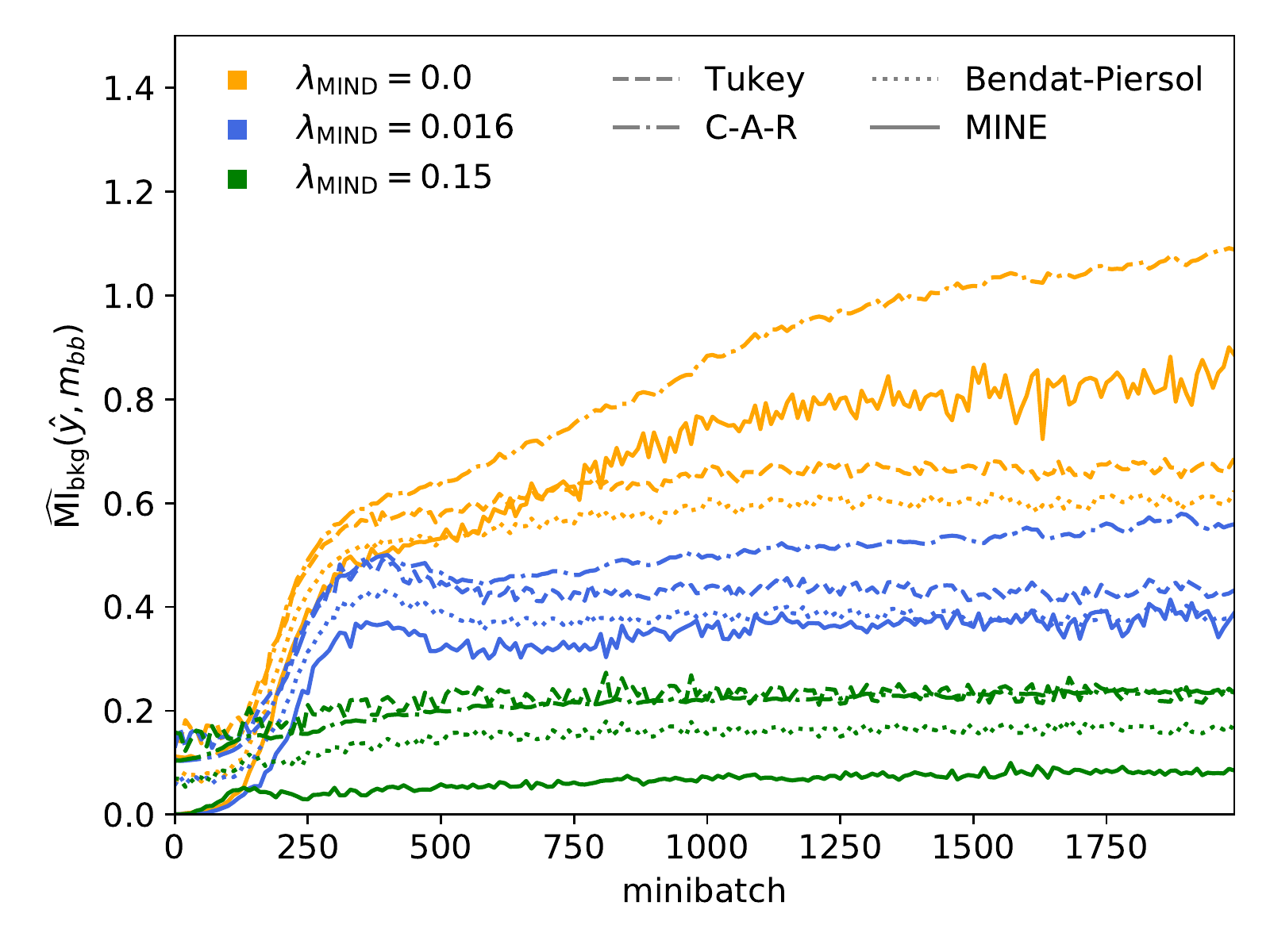}
  \caption{Evolution of various estimates of $\mathrm{MI}_{\mathrm{bkg}}(\hat{y}, m_{bb})$ with the number of minibatches seen during training with MIND, for several values of $\lambda_{\mathrm{MIND}}$. The MINE estimate (solid line) is compared with several histogram-based estimators, following the binning strategies proposed by Bendat and Piersol \cite{bendat_piersol} (dotted line), Tukey and Mosteller \cite{tukey} (dashed line) and the adaptive binning method suggested by Cellucci, Albano and Rapp (C-A-R, dash-dotted line) \cite{cellucci_binning_heuristics}.}
  \label{fig:mi_comparison}
\end{figure*}

To demonstrate that the MINE estimator introduced in Section \ref{pca} gives a reasonable assessment of MI in our setting, we compare it to a number of binned MI estimators. Figure \ref{fig:mi_comparison} shows the evolution of various estimates of the mutual information on background events, $\mathrm{MI}_{\mathrm{bkg}}(\hat{y}, m_{bb})$, for several values of $\lambda$ as the adversarial training progresses. These estimates are obtained from 100k (weighted) events. All methods show the same trend of MI and clearly indicate a reduction in its final value as $\lambda$ increases. For an untrained classifier network with randomly initialised weights, $\mathrm{MI}_{\mathrm{bkg}}(\hat{y}, m_{bb})$ is very close to zero. While MINE correctly captures this effect, all binned methods suffer from bias and indicate a nonzero value of MI even for independent variables.

\begin{figure}[htp]
  \centering
  \hspace*{-0.9cm}
  \subfloat[][]{\includegraphics[width=0.55\textwidth]{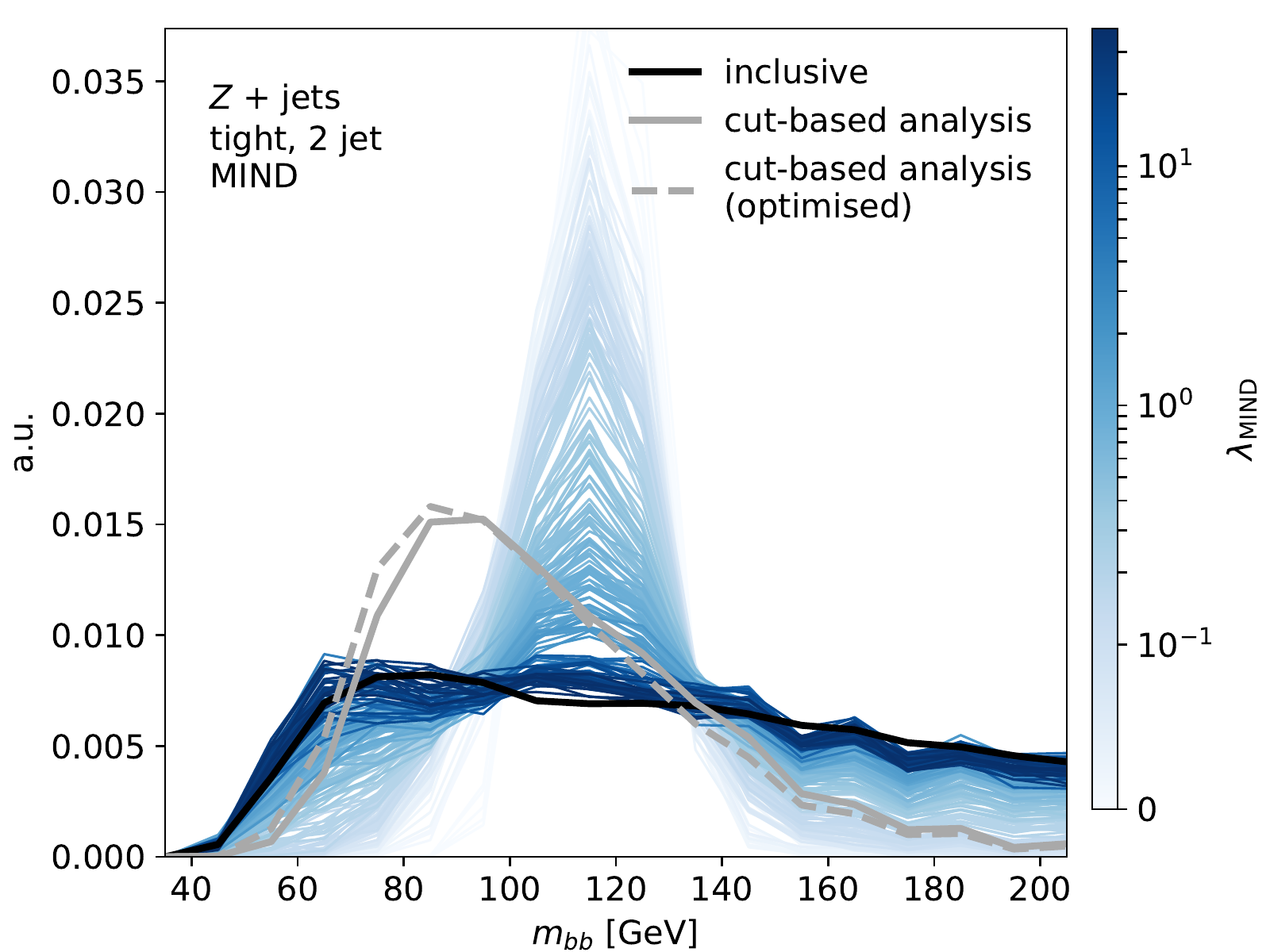}\label{Zjets_MIND}}
  \subfloat[][]{\includegraphics[width=0.55\textwidth]{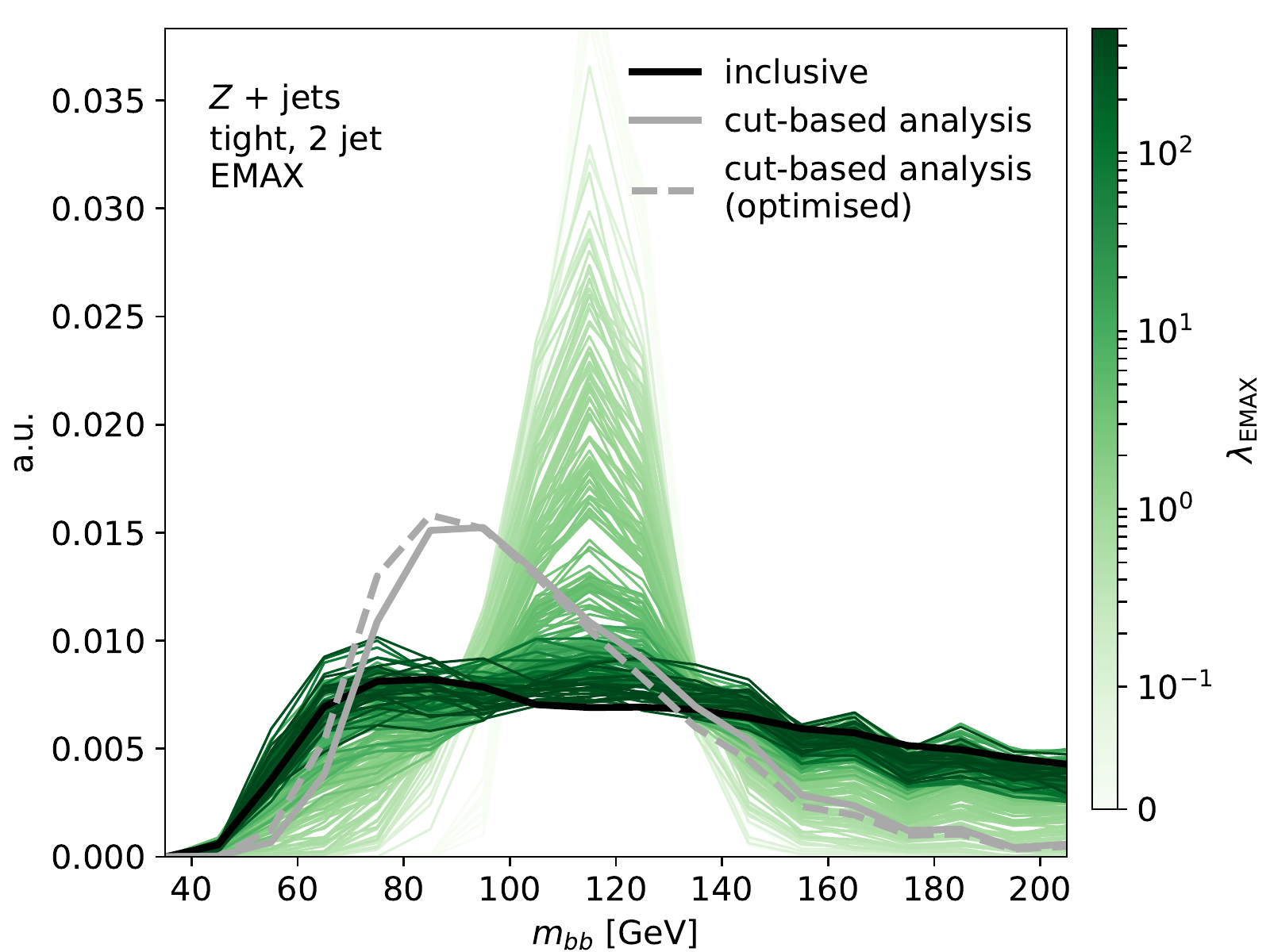}\label{Zjets_EMAX}}\\
  \hspace*{-0.9cm}
  \subfloat[][]{\includegraphics[width=0.55\textwidth]{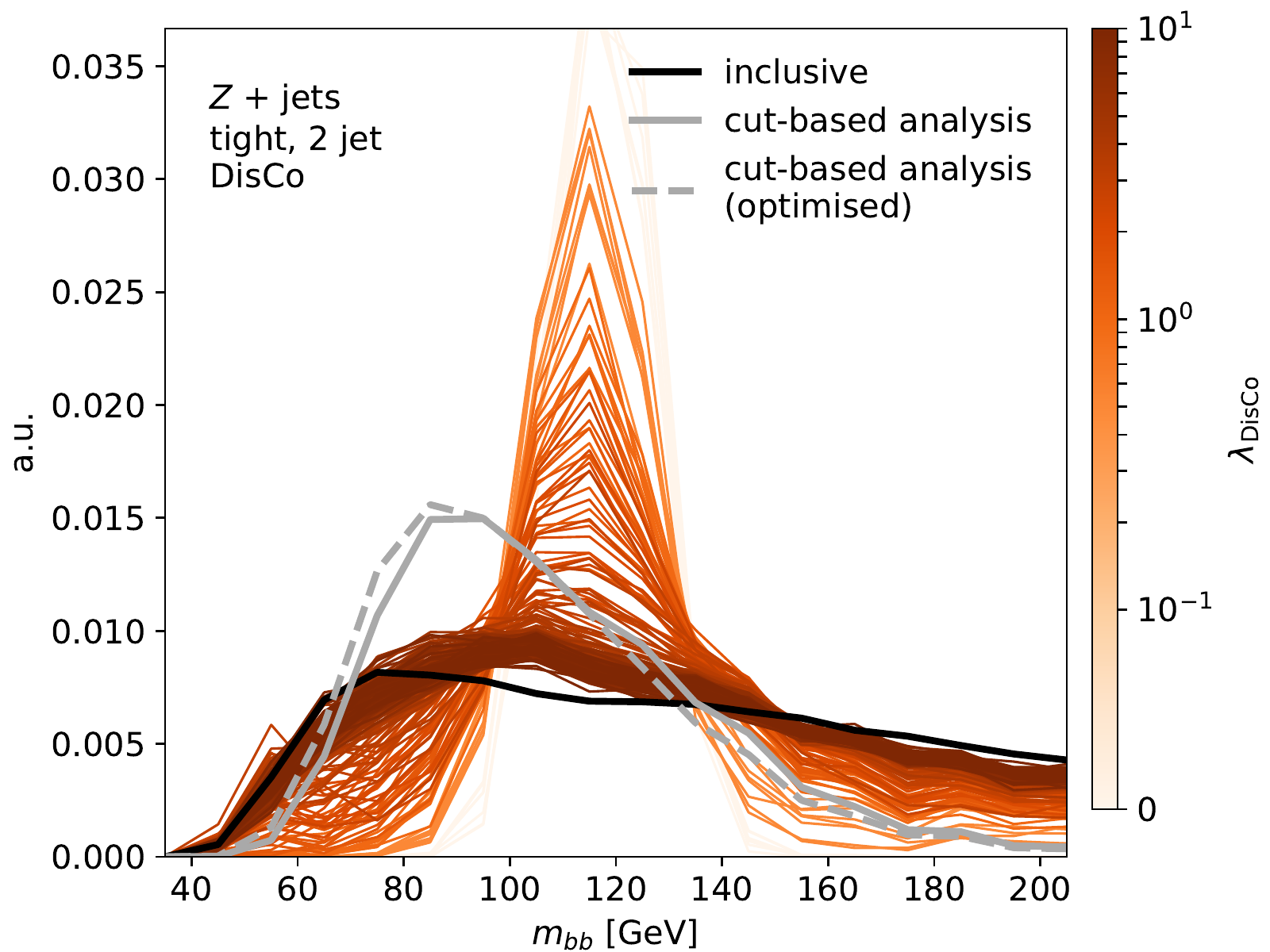}\label{Zjets_DisCo}}
  \caption{Invariant mass distributions produced by $Z$+jets events in the two-jet tight DCA categories, for MIND in \protect\subref{Zjets_MIND}, EMAX in \protect\subref{Zjets_EMAX} and DisCo in \protect\subref{Zjets_DisCo}. The shade of the lines corresponds to the value of $\lambda$ used in the training. The
    thick black line shows the inclusive distribution, i.e.~the one obtained for the corresponding jet multiplicity, but without any further selections applied. The grey lines correspond to the distributions obtained for the equivalent CBA high-$\met$ signal region. Distributions are shown both for the original (solid) and optimised (dashed) cut values.}
  \label{fig:lambda}
\end{figure}

The effect of $\lambda$ on the classifier response is illustrated in Figure \ref{fig:lambda}. It shows the invariant
mass distributions generated by the $Z$+jets background in the 2-jet tight signal region of the DCA for all three considered methods. For comparison, the distributions achieved in the equivalent CBA signal region are shown as well, as are the inclusive mass distributions, i.e.~those for the corresponding jet multiplicity, but without any selection applied.
Small values of $\lambda$ (corresponding to a traditional classifier) result in events being selected preferentially around the Higgs boson mass peak, considerably deforming the $\mbb$ distribution. Classifiers trained with a larger value of $\lambda$ select events more uniformly across the invariant mass range. Even though the shape of the total background is preserved to a good degree (see below), residual distortions to individual background components are not explicitly captured by our training setup and remain visible.

\begin{figure}[p]
  \centering
  \includegraphics[width=\textwidth]{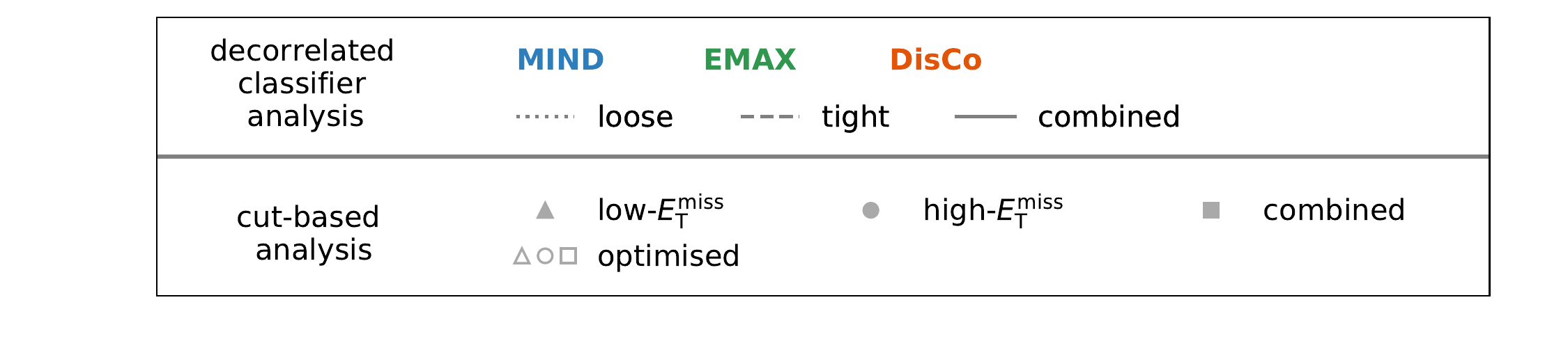}\\
  \includegraphics[width=\textwidth]{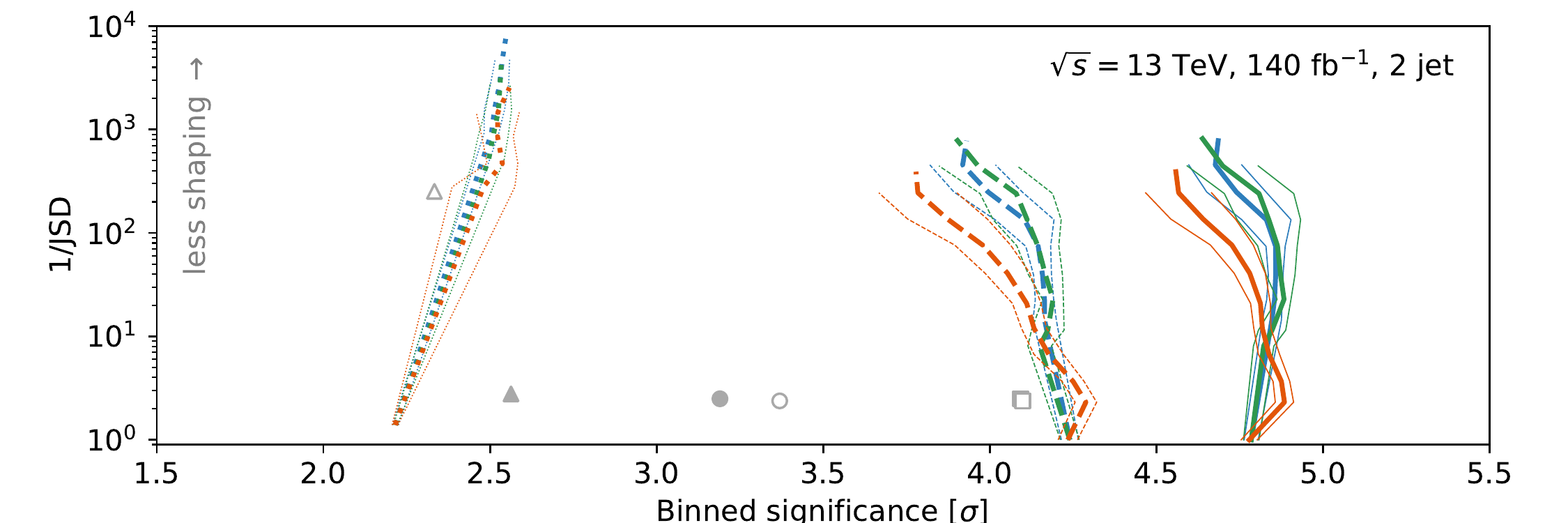}\\\vspace{0.4cm}
  \includegraphics[width=\textwidth]{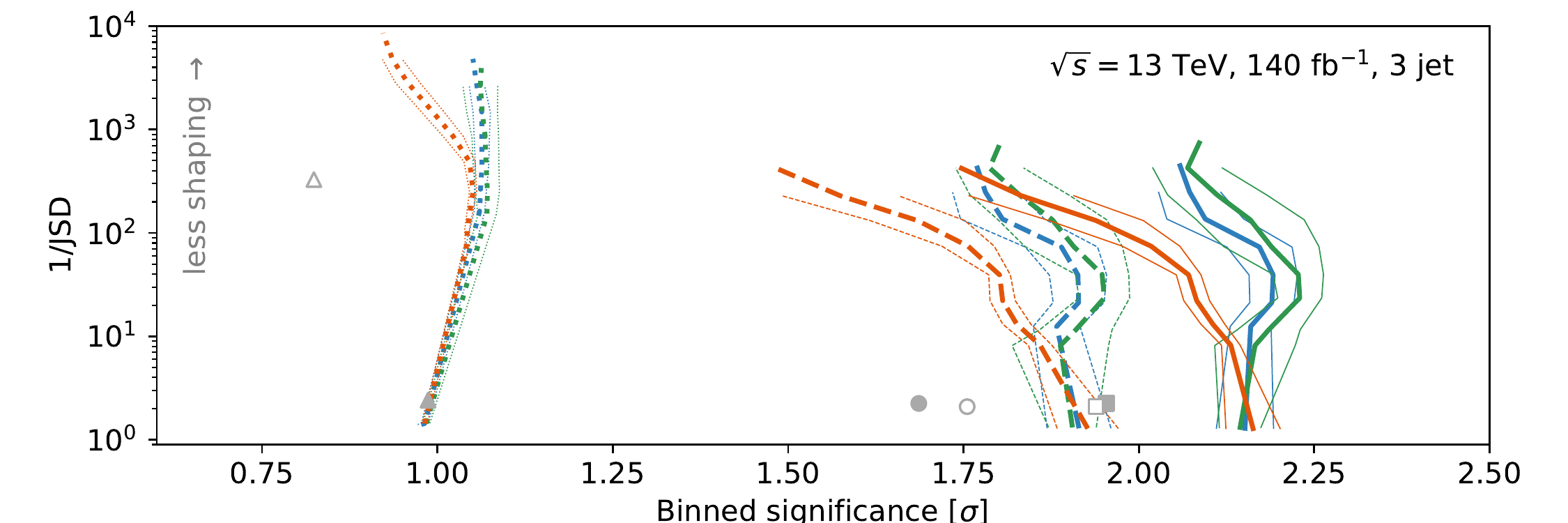}\\\vspace{0.4cm}
  \caption{Comparison of performance achieved by various DCA methods and the CBA in the two-jet (top) and three-jet (bottom) categories. The horizontal axis shows the
    binned significance estimate. The vertical axis shows the inverse of the Jensen-Shannon divergence ($1/\text{JSD}$) between the $\mbb$ distributions of the combined background in individual analysis categories and the inclusive case. For the combination of two categories, $1/\text{JSD}$ between the background distribution in the \emph{tight}
    (high-$\met$) category and the inclusive distribution is plotted for DCA (CBA). The thick lines indicate the average binned significances achieved at a certain level of $1/\text{JSD}$ and the thin lines correspond to their sample standard deviations. For each method, around 500 models are trained with various values of $\lambda$.}
  \label{fig:scatter}
\end{figure}

Figure \ref{fig:scatter} compares the various DCA and CBA event classification schemes, separately for the two-jet (top) and three-jet (bottom)
categories. For each jet multiplicity, the achieved performance is shown individually for each signal region, as well as for their combination.
The horizontal axis shows the estimate \cite{Cowan:2010js} of the binned discovery significance 
\begin{equation}
\sigma = \sqrt{2 \sum_{i} \left[(S_i + B_i)\log\left(1+\frac{S_i}{B_i}\right) - S_i \right] },
\end{equation}
where $S_i$ ($B_i$) is the number of signal (background) events in bin $i$ and the sum runs over all the bins in the $\mbb$ distribution.
The vertical axis shows the inverse of the Jensen-Shannon divergence (JSD) between the $\mbb$ distributions for the combined background in each of the analysis categories and the inclusive background distribution. The Jensen-Shannon divergence is a measure of similarity between two distributions $P$ and $Q$, and is a symmetrised version of the Kullback-Leibler divergence,
\begin{equation}
\text{JSD}(P||Q) = \frac{1}{2} \left[ D_\text{KL}(P||M) + D_\text{KL} (Q || M)\right],
\end{equation}
where $M = \frac{1}{2}(P+Q)$.

By virtue of Equation \ref{independence}, this distortion measure directly indicates the extent to which $\mbb$ and $\hat{y}$ are statistically independent random variables. Classifiers trained with low $\lambda$ heavily distort the $m_{bb}$ distributions, corresponding to low values of $1/\text{JSD}$. Larger values of $1/\text{JSD}$ imply a higher degree of independence, i.e.~correspond to less distortion of the $\mbb$ spectrum in the signal regions. 

With respect to these metrics, all three evaluated DCA methods clearly outperform CBA and achieve higher significances, even in situations where the distortion of the mass distributions is well below the levels achieved by CBA. MIND and EMAX show identical performance and are able to retain significant classification performance as the shaping is reduced through the adversarial training. In particular, the binned significances achieved by the four analysis categories remain almost constant as the level of decorrelation increases. This indicates that the classifier performance in the $m_{bb}$-range dominated by the signal remains practically unaffected by a more uniform background acceptance. DisCo shows good convergence behaviour everywhere, but leads to reduced binned significances at high values of $\lambda$, in particular in the 3-jet channel\footnote{In terms of the DisCo-loss, the solutions found by MIND perform worse than those found by DisCo itself, i.e.~achieve higher loss values. This points to practical differences in the objectives specified by DisCo and MIND.}.

We stress that the binned significance measures the analysis performance when uncertainties on the background are absent. In practice, this will not be the case, and in particular the background normalisation will generally be extracted together with the signal yield. As a more realistic measure for the overall analysis performance, a combined Asimov fit to the $\mbb$ distributions in all four signal regions is performed. All signal- and background components are scaled by separate, unconstrained normalisation factors that are determined simultaneously by the fit. The signal discovery significance obtained in this way gives a combined performance measure, taking into account also the distortion of the mass distribution produced by the event classification.
Effects coming from systematic uncertainties are not included in the fit. Extracting realistic estimates of their relative impact on the various analysis methods necessitates a more realistic dataset and the choice of a specific systematics model, the study of which we leave to future work.

\begin{figure*}[htp]
  \centering
  \includegraphics[width=0.72\textwidth]{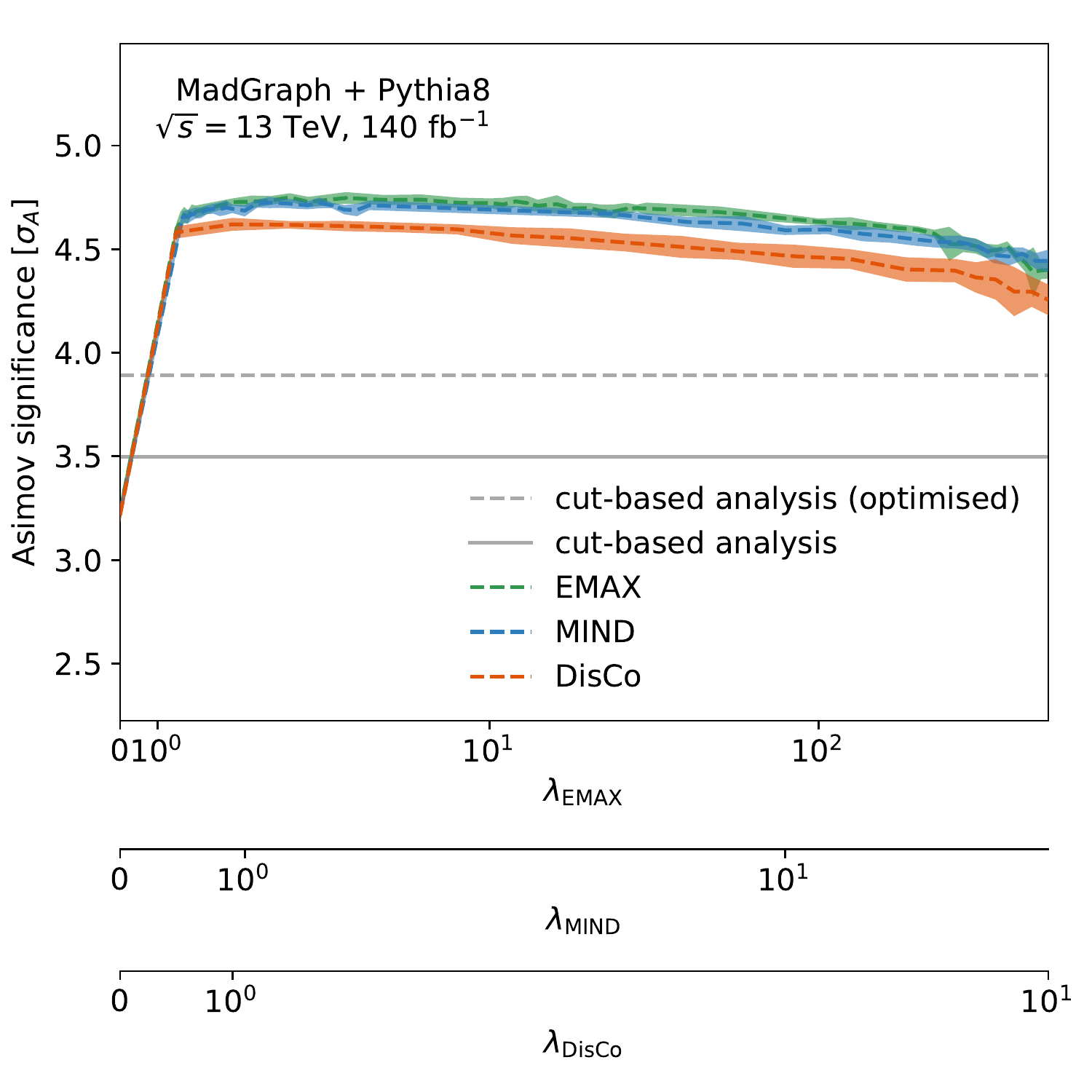}
  \caption{Asimov discovery significance $\sigma_A$ for $VH\rightarrow b\bar{b}$ in the 0-lepton channel as a function of the hyperparameter $\lambda$, separately for the three investigated DCA strategies. The significance expected for the CBA using the original (optimised) cut values is indicated by the grey (dashed) line. For each value of $\lambda$, the dashed line shows the mean significance, while the colored uncertainty band corresponds to the sample standard deviation computed from 10 independent training runs. The axis scaling is chosen to highlight the similar behaviour of all three DCA methods.}
  \label{fig:asimov}
\end{figure*}

Figure \ref{fig:asimov} shows the Asimov discovery significance obtained from this fit for the signal process $VH \rightarrow b\bar{b}$. It contrasts the results for the CBA with the significances obtained for the DCA as a function of $\lambda$, for MIND, EMAX and DisCo. For $\lambda = 0$, all three methods become equivalent and correspond to a regular classifier without decorrelation. In this case, the $m_{bb}$ distributions of the backgrounds are heavily distorted in all four signal regions. In particular, this impacts the ability of the fit to constrain the background normalisations, leading to a severe deterioration of the extracted Asimov fit significance, which is even below that achieved by the CBA.
As $\lambda$ is increased, the shaping of the backgrounds is dramatically reduced, resulting in an enhanced performance of the DCA. Also in terms of the fit significance, MIND and EMAX are seen to achieve virtually identical performance, surpassing the sensitivity set by the CBA by at most 21\% and by about 13\% at the highest levels of decorrelation. The latter is equivalent to an increase in the size of the dataset of about 28\%. DisCo leads to a peak improvement of 18\% at intermediate levels of decorrelation, which drops to 9\% at the highest chosen values of $\lambda_{\mathrm{DisCo}}$.

\begin{figure}[!!b]
  \centering
  \subfloat[][]{\includegraphics[width=0.48\textwidth]{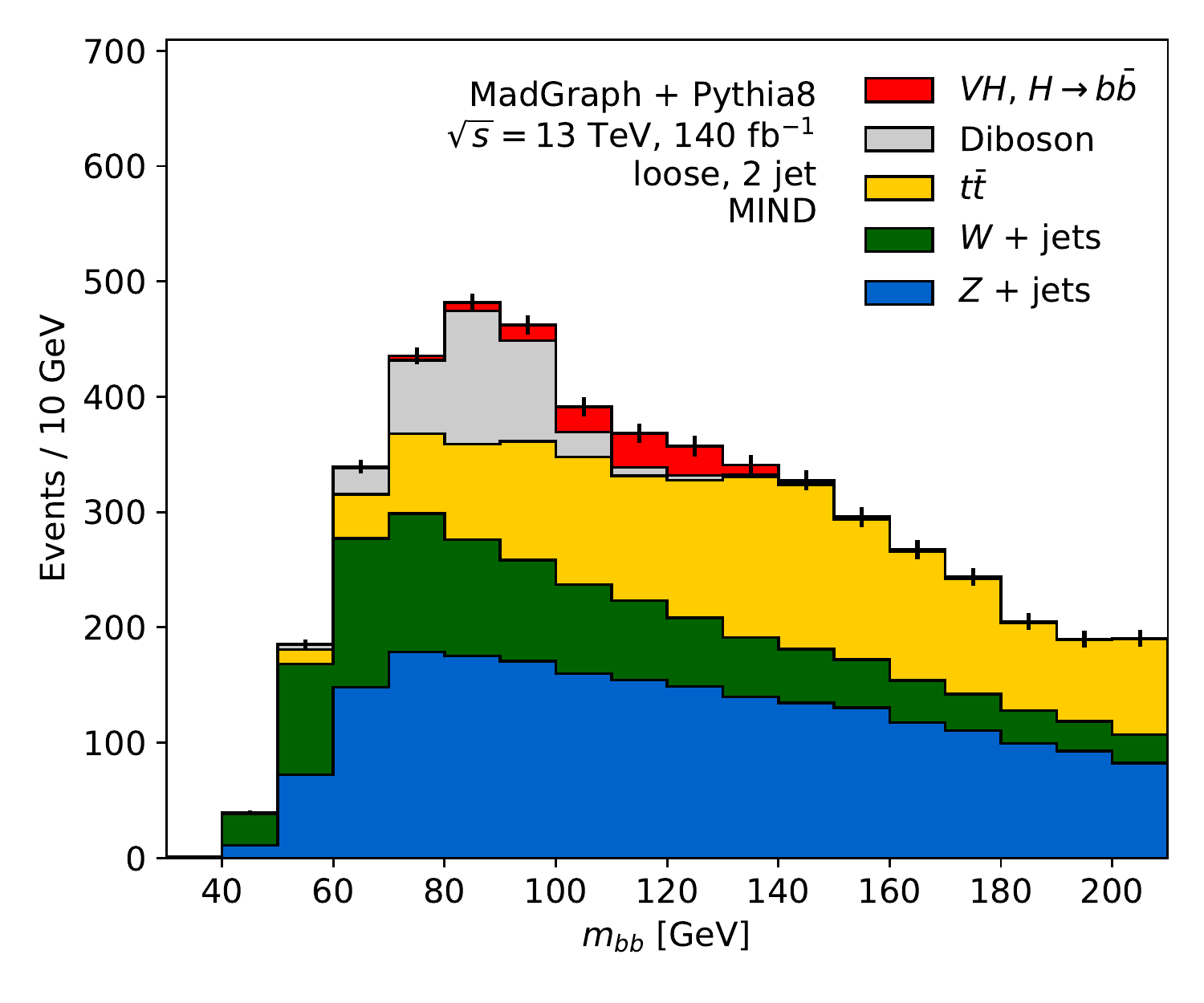}}
  \subfloat[][]{\includegraphics[width=0.48\textwidth]{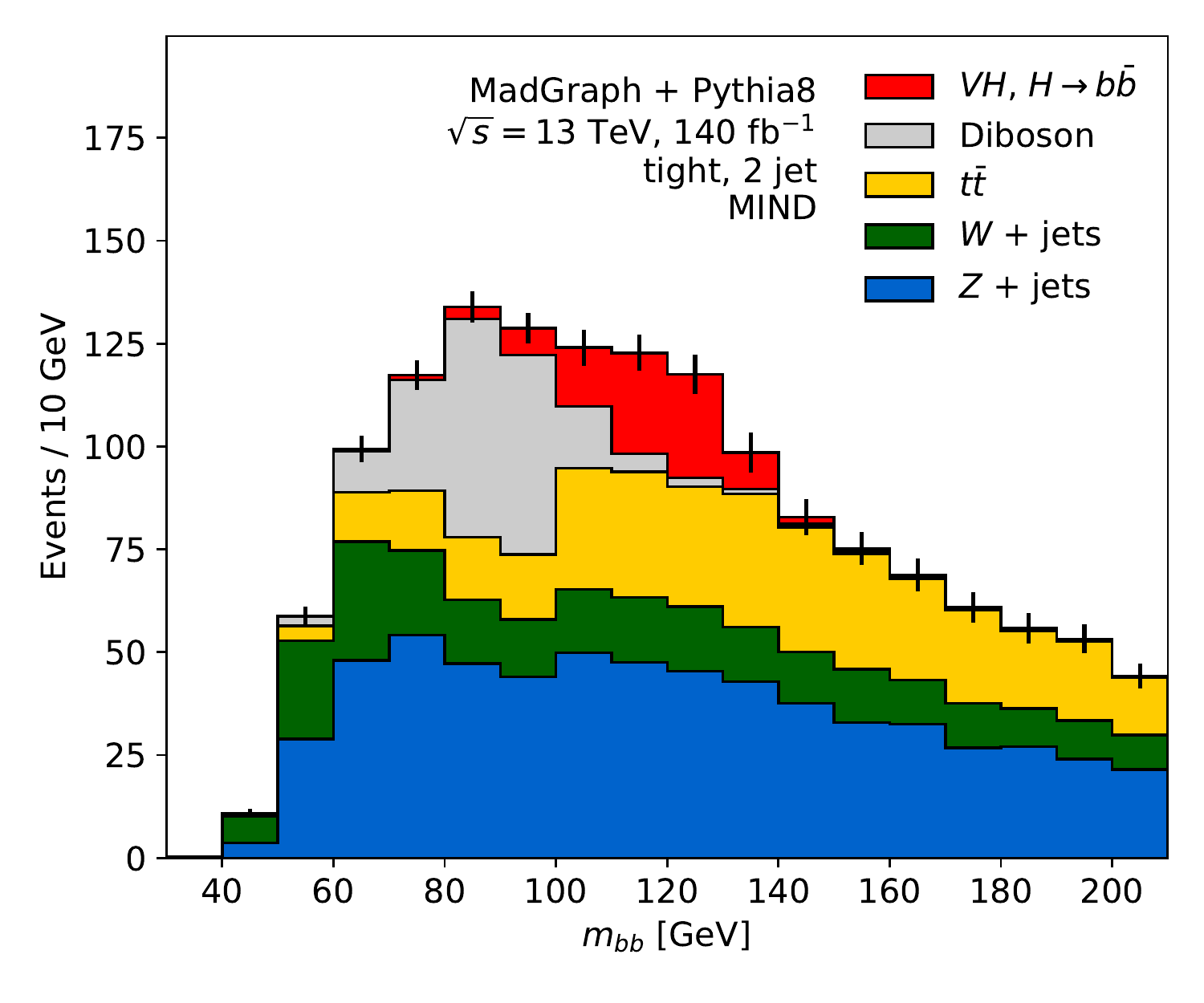}}\\
  \subfloat[][]{\includegraphics[width=0.48\textwidth]{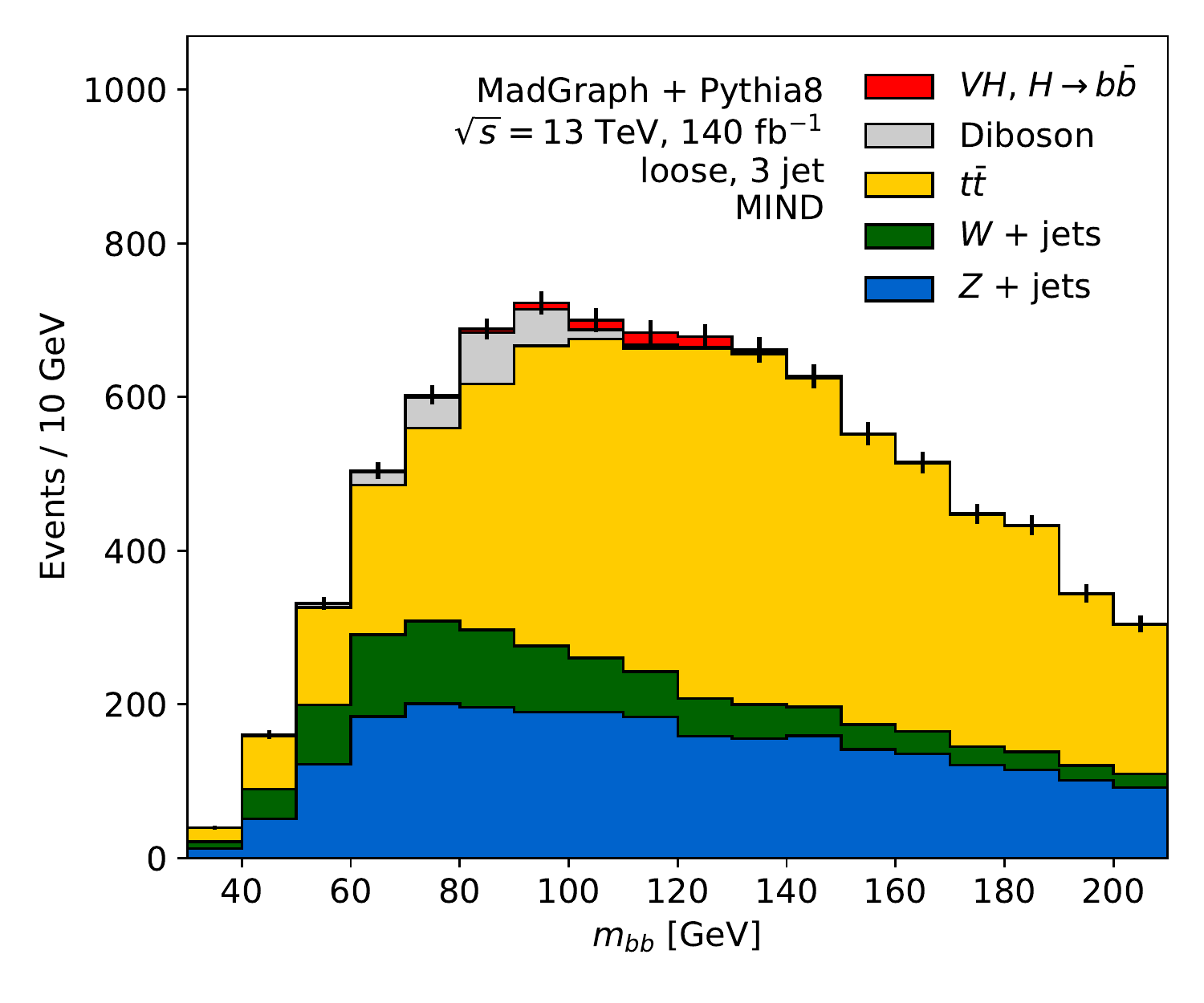}}
  \subfloat[][]{\includegraphics[width=0.48\textwidth]{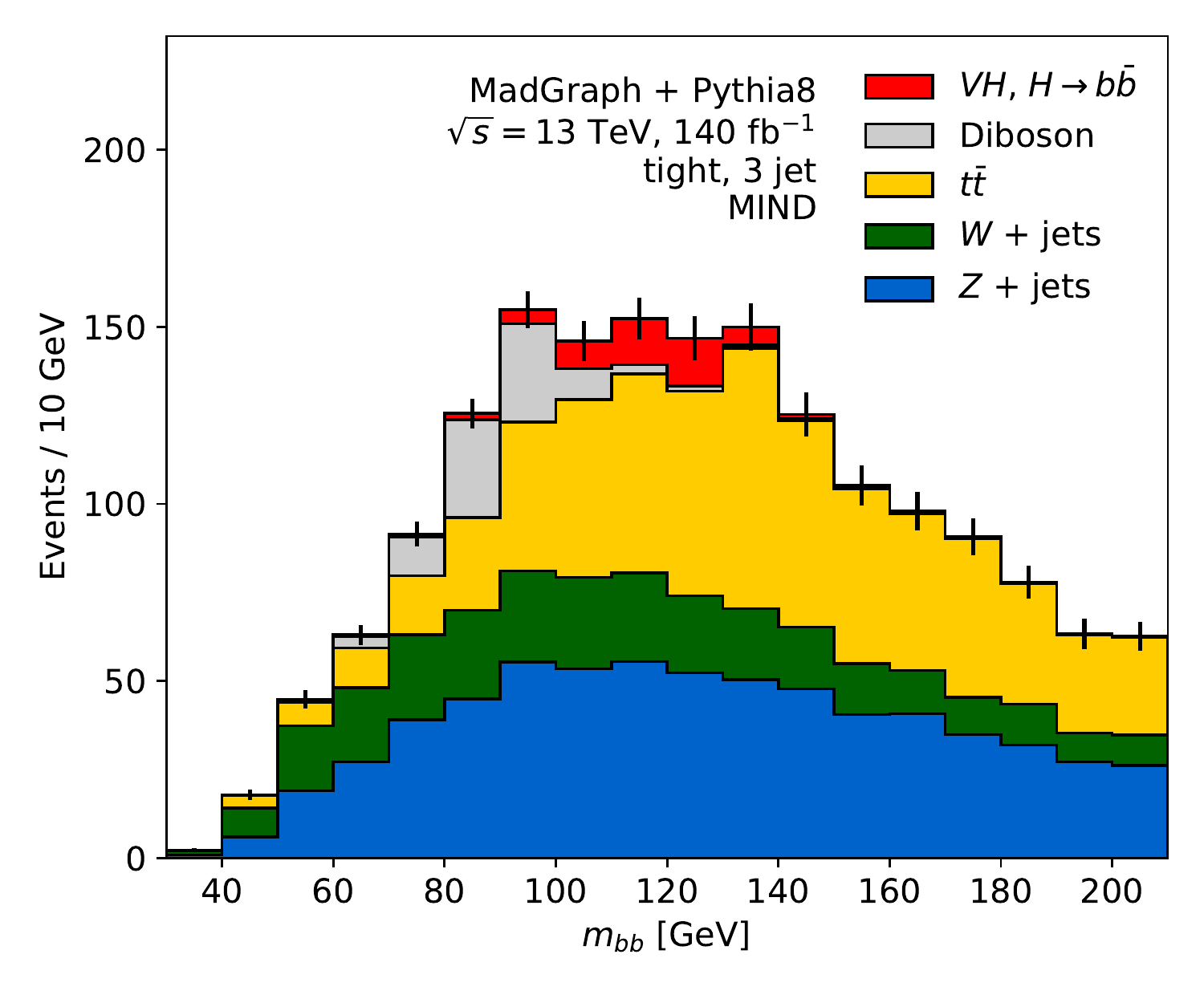}}\\
  \caption{Analysis categories for a decorrelated classifier trained with MIND and using $\lambda_{\mathrm{MIND}} = 40$. The $\mbb$ distribution is shown in the two-jet
  \textit{loose} (a) and \textit{tight} (b) categories, as well as three-jet \textit{loose} (c) and \textit{tight} (d) categories.
    Signal events are shown in red, while backgrounds are shown in blue (\textit{Z}+jets), green (\textit{W}+jets), yellow
    ($\ttbar$), and gray (diboson). Error bars show the statistical uncertainty on the total MC prediction and are dominated by the
    uncertainty on the background.}
  \label{fig:pca}
\end{figure}

To illustrate the composition of the four DCA signal regions, Figure \ref{fig:pca} shows the $\mbb$ distributions of the signal and main backgrounds for a classifier trained with $\lambda = 40$ using MIND. This particular model achieves an Asimov significance in the simultaneous fit of about $4.4\,\sigma$.

\section{Conclusions}
\label{conclusions}
Separating contributions from signal and background processes is a ubiquitous problem in the analysis of collider data. In this work, we presented a selection procedure based on a discriminating variable computed by a neural network. We introduced a new method to render this discriminant statistically independent from a set of physically important variables, whose distributions are thus unaffected by the event selection.

We demonstrated the practical feasibility of our technique in the study of the 0-lepton channel of the $VH \rightarrow b\bar{b}$ process. Constructed to preserve the spectrum of the invariant mass of the two $b$-jets, our discriminant achieves a gain in analysis sensitivity of about 21\% compared to a conventional cut-based approach and achieves state-of-the-art performance when compared to other decorrelation techniques proposed in the literature.

Based on the adversarial minimisation of mutual information, this method is expected to generalise seamlessly to situations where the preservation of multiple variables is desired. In addition, achieving independence from a set of continuous and discrete variables (e.g.~discrete nuisance parameters) can also be considered, a case whose study we leave for future investigation.

\acknowledgments

The authors are grateful to Gregor Kasieczka and David Shih for help with DisCo and for pointing out a problem with the implementation of the original decorrelation approach. Credit also goes to Kyle Cranmer, Gilles Louppe, Johnny Raine, David Rousseau, and Justin Tan for useful discussions, and to Anastasia Kotsokechagia for feedback on the presentation of our results. MZ and PW are supported by the Science and Technology Facilities Council (STFC) under the grant ST/N504233/1 and ST/S505638/1. In addition, PW is supported by a Buckee scholarship awarded by Merton College, Oxford.

\bibliographystyle{JHEPmod}
\bibliography{bibs}

\end{document}